# Automatic Acute Ischemic Stroke Lesion Segmentation using Semi-supervised Learning

Bin Zhao[a], Shuxue Ding[a,b] , Hong Wu[a], Guohua Liu[a], Chen Cao[c], Song Jin[c], Zhiyang Liu[a,*]

a Tianjin Key Laboratory of Optoelectronic Sensor and Sensing Network Technology, College of Electronic Information and Optical Engineering, Nankai University, Tianjin 300350, China
b School of Artificial Intelligence, Guilin University of Electronic Technology, Guilin Guangxi 541004, China
c Key Laboratory for Cerebral Artery and Neural Degeneration of Tianjin, Department of Medical Imaging, Tianjin Huanhu Hospital, Tianjin 300350, China

**ABSTRACT**
Ischemic stroke has been a common disease in the elderly population, which can cause long-term disability and even death. However, the time window for treatment of ischemic stroke in its acute stage is very short. To fast localize and quantitively evaluate the acute ischemic stroke (AIS) lesions, many deep-learning-based lesion segmentation methods have been proposed in the literature, where a deep convolutional neural network (CNN) was trained on hundreds of fully-labeled subjects with accurate annotations of AIS lesions. Such methods, however, require a large number of subjects with pixel-by-pixel labels, making it very time-consuming in data collection and annotation. Therefore, in this paper, we propose to use a large number of weakly-labeled subjects with easy-obtained slice-level labels and a few fully-labeled ones with pixel-level annotations, and propose a semi-supervised learning method. In particular, a double-path classification network (DPC-Net) was proposed and trained using the weakly-labeled subjects to detect the suspicious AIS lesions. A K-Means algorithm was used on the DWIs to identify the potential AIS lesions due to the a priori knowledge that the AIS lesions appear as hyperintense. Finally, a region-growing algorithm combines the outputs of the DPC-Net and the K-Means to obtain the precise lesion segmentation. By using 460 weakly-labeled subjects and 5 fully-labeled subjects to train and fine-tune the proposed method, our proposed method achieves a mean dice coefficient of 0.642, and a lesion-wise F1 score of 0.822 on a clinical dataset with 150 subjects.

## 1. Introduction

Stroke has been one of the most common causes of death and long-term disability worldwide [1], which brings tremendous pain and financial burden to patients. In general, stroke can be categorized as ischemia and hemorrhage according to the types of cerebrovascular accidents, where ischemic stroke accounts for 87% [2]. As the ischemic stroke may lead to invertible damage on brain tissues, in clinical practice, it is of paramount importance to quickly diagnose and quantitively evaluate in the acute stage to improve the treatment outcome.

In diagnosing of ischemic strokes, magnetic resonance imaging (MRI) serves as the modality of choice for clinical evaluation. The diffusion weighted images (DWIs) and the apparent diffusion coefficient (ADC) maps derived from multiple DWIs with different b-values have been shown to be sensitive in diagnosing acute ischemic stroke (AIS). In particular, the AIS lesions appear as hyperintense on the DWIs and hypointense on the ADC maps [3]. Fig. 1 presents some examples of AIS lesions. The regions identified by the red arrows are AIS lesions. The regions identified by the yellow arrows, although also shown as hyperintense on the DWIs, are non-lesion regions. In fact, such hyperintensive regions are the magnetic susceptibility artifacts. That is to say, despite that they appear as hyperintense on the DWIs, there is no abnormality on the ADC maps. Therefore, to correctly identify the lesions, it is important to jointly consider both DWIs and ADC maps to extract the semantic information.

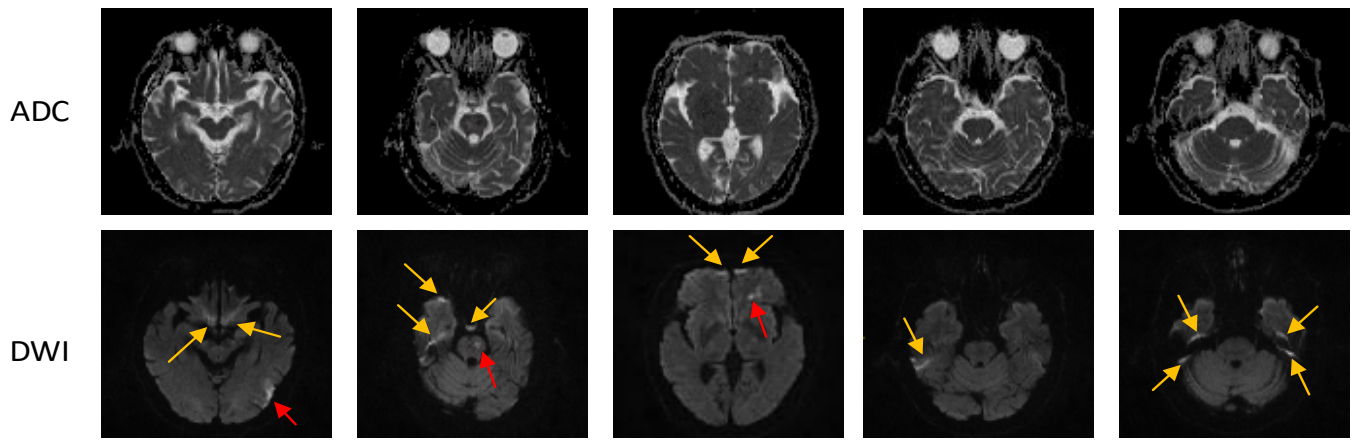

Fig. 1. Challenge examples in AIS segmentation. The first row show ADC slices and the second row shows their corresponding DWI slices. The yellow arrows identify the hyperintense due to magnetic susceptibility artifacts, and the red arrows identify the hyperintense that are true AIS lesions. Best viewed in color.

Recently, convolutional neural network (CNN) based methods have presented tremendous ability in image classification and semantic segmentation on medical image processing [4-6]. Different from conventional image processing methods using handy-crafted image features, the CNN-based methods extract features from manually labeled data by itself. By training CNNs on a massive amount of labeled images, the CNN-based methods have achieved promising results on various organ and lesion segmentation tasks [7-10]. In order to make use of contextual information in volumetric data, in [11], a 3D CNN was trained to automatically localize and delineate the abdominal organs of interest with a probability prediction map, and then a time-implicit multi-phase level-set algorithm was utilized for the refinement of the multi-organ segmentation. Another 3D CNN-based method using fully convolutional DenseNet [12] was proposed for automatic segmentation of AIS [13], which achieved high performance on their test set.

* Corresponding author at: Tianjin Key Laboratory of Optoelectronic Sensor and Sensing Network Technology, College of Electronic Information and Optical Engineering, Nankai University, Tianjin, china
*E-mail address:* liuzhiyang@nankai.edu.cn (Z. Liu).

Typically in a CNN, millions of parameters have to be tuned, and therefore a massive amount of images with accurate annotations are required. For instance, [8] used 165 fully-labeled subjects for organ segmentation, and [13] collected 152 fully-labeled subjects for lesion segmentation. Different from the images in the ImageNet [14] and COCO [15] dataset which can be easily obtained from the Internet and labeled by the ordinary people, the medical images have to be acquired by special equipment, and many well-trained clinicians are further required to precisely annotate the labels. More importantly, to perform image segmentation, most methods require pixel-by-pixel annotations to train the CNN [6, 11, 16-23]. For instance, in the AIS lesion segmentation task, most methods [5, 13, 24] required subjects with pixel-wise labels as shown in Fig. 2, where each pixel was annotated as normal or lesion. Obviously, annotating the pixel-level labels are labor-intensive and time-consuming, making it even difficult to establish a large dataset. This motivates us to develop a segmentation method by using much simpler annotations. One of the simpler annotations is just annotating whether each slice incorporates lesions or not, as shown in Fig. 3. Hereafter, we term this simpler annotation as weak annotation, and the data samples with weak annotations as weakly-labeled. Since the clinicians only require annotating whether each slice includes AIS lesions or not, the annotation cost can be significantly reduced, making it easier to collect a large amount of labeled data samples.

With weakly-labeled data subjects, the CNN can generate a class activation map (CAM) for the AIS lesions [25]. The CAMs, however, were far from accurate segmentation, as they were in fact obtained from the weighted sum of low-resolution feature maps of a convolution layer. As shown in [25], despite that the lesion-wise detection rate is high, the CAMs only cover the most prominent part of the lesion. In fact, merely using weakly-supervised learning would inevitably lead to an underestimated segmentation for a large lesion, and an overestimated one for a small lesion. As we will show in this paper, by fusing multi-resolution feature maps, the proposed double-path classification network (DPC-Net) is able to output a suitable estimation for AIS lesion segmentation.

To generate a precise segmentation, we further propose to use a small amount of fully-labeled subjects to facilitate the segmentation. Despite that such semi-supervised learning methods have been adopted in image segmentation tasks by using a mixed dataset with many unlabeled samples and a few fully-labeled samples [26-28], most methods propose to adopt a self-training or co-training strategy, where the unlabeled samples were automatically annotated and added to the labeled set. Note that the generated labels will include errors, and some errors will be even strengthened during the iterative training process, leading to a risk of performing worse than a supervised approach. To solve this problem, we propose to utilize the a priori knowledge, and use the fully-labeled subjects to fine tune a machine learning method with much fewer parameters.

In particular, we propose to use the K-Means algorithm on the DWIs due to the a priori knowledge that the AIS lesions appear as hyperintense on the DWI, so that a number of candidate AIS lesion regions can be obtained. Then a region growing algorithm is adopted to combine the semantic information from the DPC-Net and the textural and spatial information from the K-Means. Different from the conventional region-growing method, the algorithm adopted in this paper can automatically find an initial point from the DPC-Net rather than manually assign one. By training on 460 weakly-labeled and 5 fully-labeled subjects, the proposed method is able to achieves a mean dice coefficient of 0.642, and a lesion-wise F1 score of 0.822 on a clinical dataset with 150 subjects.

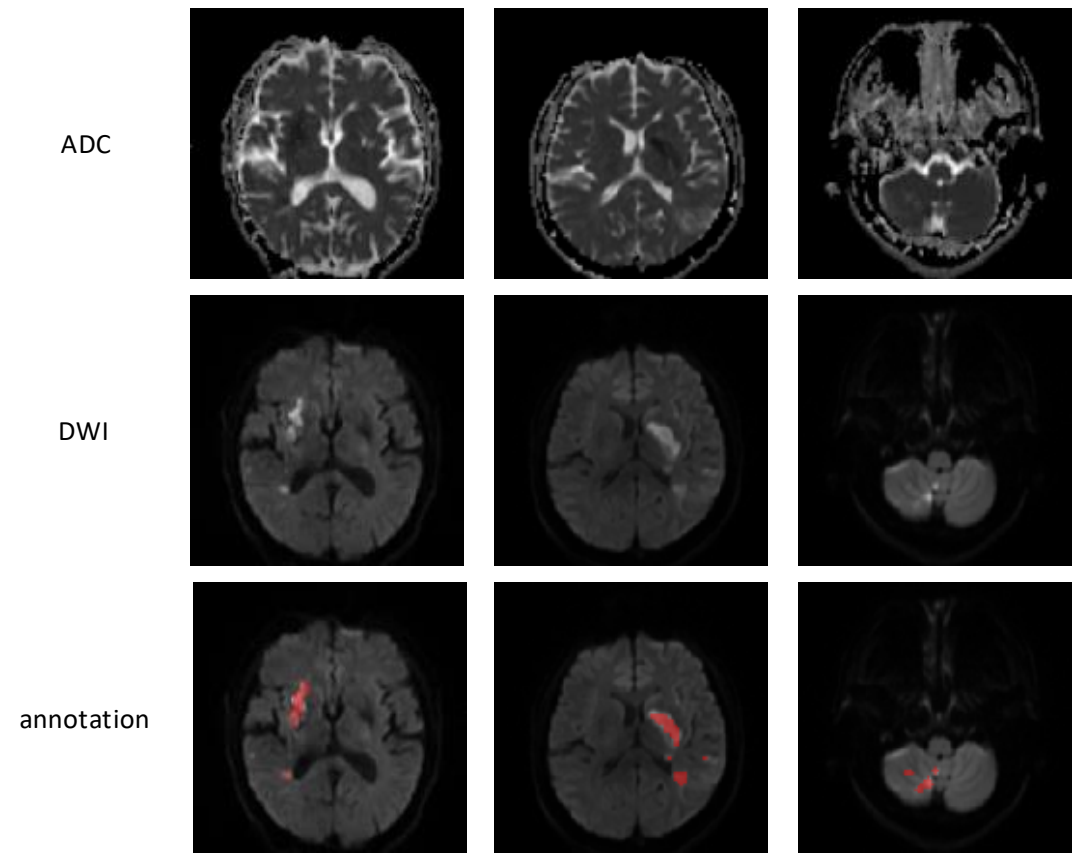

Fig. 2. Examples of the fully-labeled subjects. The first two rows show ADC slices and their corresponding DWI slices. The third row shows the annotations. Best viewed in color.

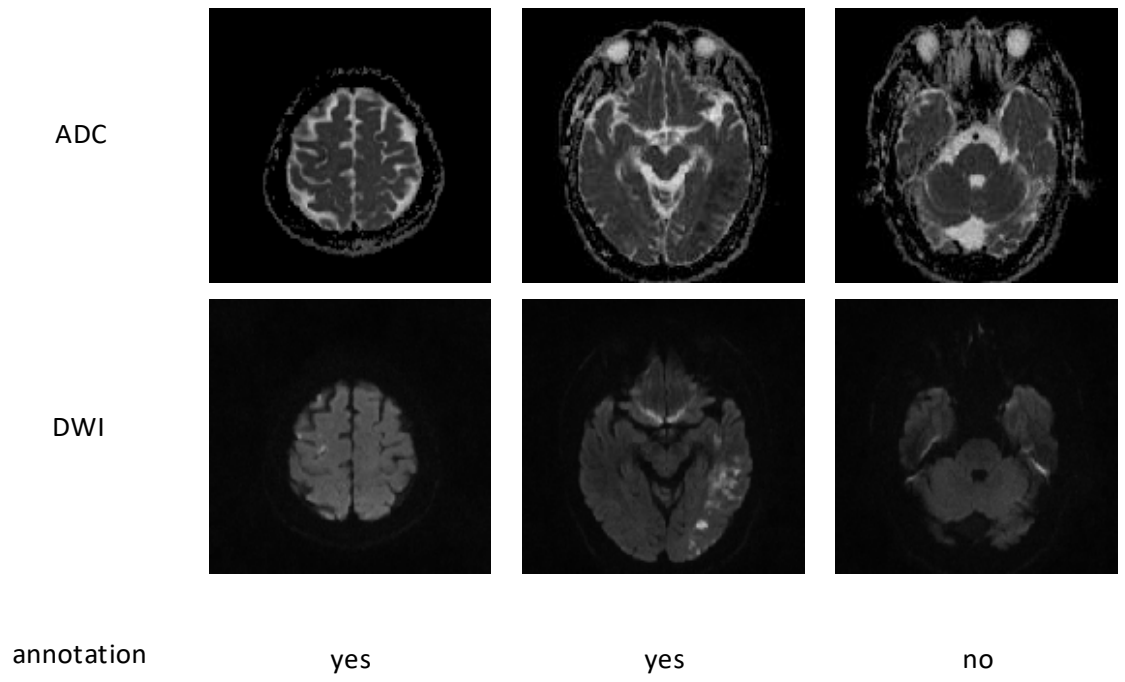

Fig. 3. Examples of the weakly-labeled subjects. The first two rows show ADC slices and their corresponding DWI slices. The third row shows the annotations, "yes" indicates that the slice has lesion and "no" indicates the opposite

## 2. Related work

There have been extensive efforts on automatic segmentation of ischemic stroke lesion recently. We roughly divided these methods into two categories, conventional methods and CNN-based methods, according to whether handy-crafted features are required. In the conventional methods, by defining some specific features on the MRIs, the ischemic stroke lesions can be identified by conventional image processing techniques or using machine-learning algorithms such as the random forest or the support vector machine (SVM). For instance, a gravitational histogram optimization based method was proposed for ischemic stroke lesion segmentation using DWI [29]. To reduce the false positive rate, [30] further proposed to use multimodal MRIs to extract features and identify the lesions using random forest. In [31], a stroke lesion segmentation method based on local features extracted from multimodal MRI, and a support vector machine (SVM) classifier was further trained to segment the lesions. A random-forest-based method was further used to identify the sub-acute ischemic stroke lesions [32], and achieves a top-ranking result in Ischemic Stroke Lesion Segmentation (ISLES) challenge in 2015 [33]. However, the performances of conventional methods were still not good enough as they are heavily dependent on handy-crafted features, which is understandable by the results of the recent challenge in ISLES 2015 [33].

At the same time, CNN-based methods have emerged as a powerful alternative for automatic segmentation of ischemic stroke lesion. For instance, most top-ranking methods in ISLES 2015 were based on CNNs, and the winner, DeepMedic [34], was a 3D CNN-based method, which achieved a Dice coefficient (DC) of 0.59 on test set. Some other successful methods [13, 19] based on CNNs in ISLES 2015 were also derived from generic CNNs architecture. Nevertheless, sub-acute ischemic stroke lesions have different imaging modality and lesion features from AIS lesions, so that the methods can not be straightforwardly generalized to be used for AIS lesions. It is thus necessary to explore the methods for segmentation of AIS lesion.

Due to the lack of public dataset on the AIS lesion segmentation task, the AIS lesion segmentation methods were evaluated on institutional data were reported in the literature. A framework [24] with the combination of EDD Net and multiscale convolutional label evaluation net (MUSCLE Net) achieved a DC of 0.67 on test set by using DWIs. In order to fully utilize the MRI information, Res-CNN [5] carried out segmenting AIS lesion in multi-modality MRI. To take advantage of contextual information in volumetric data, a method based on 3D fully CNNs was proposed for AIS lesion segmentation [13], which is shown to be superior in terms of precision rate. However, these methods required hundreds of high quality fully-labeled subjects, which are very time-costly to obtain.

To reduce the burden in annotation, some weakly-supervised and semi-supervised segmentation methods were proposed. In weakly-supervised learning methods, the clinicians only need to annotate whether an image contains the target tissue or not. For instance, in [35, 36], the cancer cells were segmented by using the labels that indicates whether an histopathology image contains cancer cells or not. When adopted in AIS segmentation, the weakly-supervised method presented high lesion-wise detection rate [25]. However, due to the lack of information on lesion positions, the CNN tends to only identify the most prominent feature that helps in classification, leading to an underestimated segmentation size when the lesion is large. For small lesions, as the segmentation results were in fact upsampled from the low-resolution feature maps in the final convolution layer, it tends to overestimate the lesion size. It implies that to accurately segment the lesions, some full annotations should be used.

The semi-supervised learning was conventionally referred to the case where the training data samples were mixed by both fully-labeled and unlabeled subjects. In medical image segmentation tasks, the network usually first learnt from the fully-labeled subjects to generate fake labels for those without full labels, and then update the network parameters based on the whole training set with both real and fake labels [26-28]. Such methods, however, risk in error propagation, as all knowledge of the target distribution came from the limited number of fully-labeled subjects.

To avoid this problem, in this paper, we propose to use a semi-supervised learning method with a few fully-labeled and many weakly-labeled subjects, where the weakly-labeled ones were used to train a CNN to extract semantic features and the fully-labeled ones were adopted to fine tune the segmentation results.

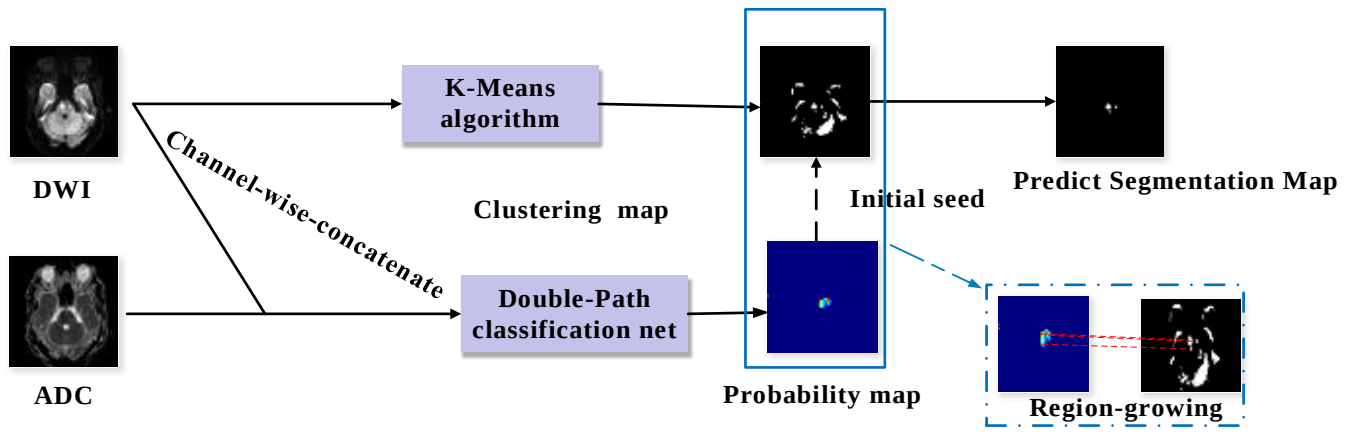

Fig. 4. Architecture of the proposed semi-supervised learning method for lesion segmentation. Best viewed in color.

## 3. Methods

In this section, we present a semi-supervised deep learning method for AIS lesion segmentation on multi-modal MR images, and the whole pipeline is presented in Fig. 4. In particular, our proposed method includes two pathways. In the first pathway, we propose a double-path classification network (DPC-Net) to extract semantic information. It is able to generate probability maps for AIS lesions with higher spatial accuracy than the CAMs in the conventional weakly-supervised learning method. In the other pathway, a K-Means clustering algorithm is incorporated to obtain the lesion segmentation results, thanks to the a priori knowledge that the AIS lesions appear as hyperintense on the DWIs. Then, a region-growing algorithm is adopted to combine the results of the two pathways and generate the final segmentation result. Compared to other semi-supervised learning based methods, the proposed method does not generate any fake labels for the weakly-labeled subjects, and therefore prevent propagating the errors from a network that overfits on such a small number of fully-labeled subjects.

In the following subsections, we will introduce the proposed semi-supervised AIS lesion segmentation method in detail.

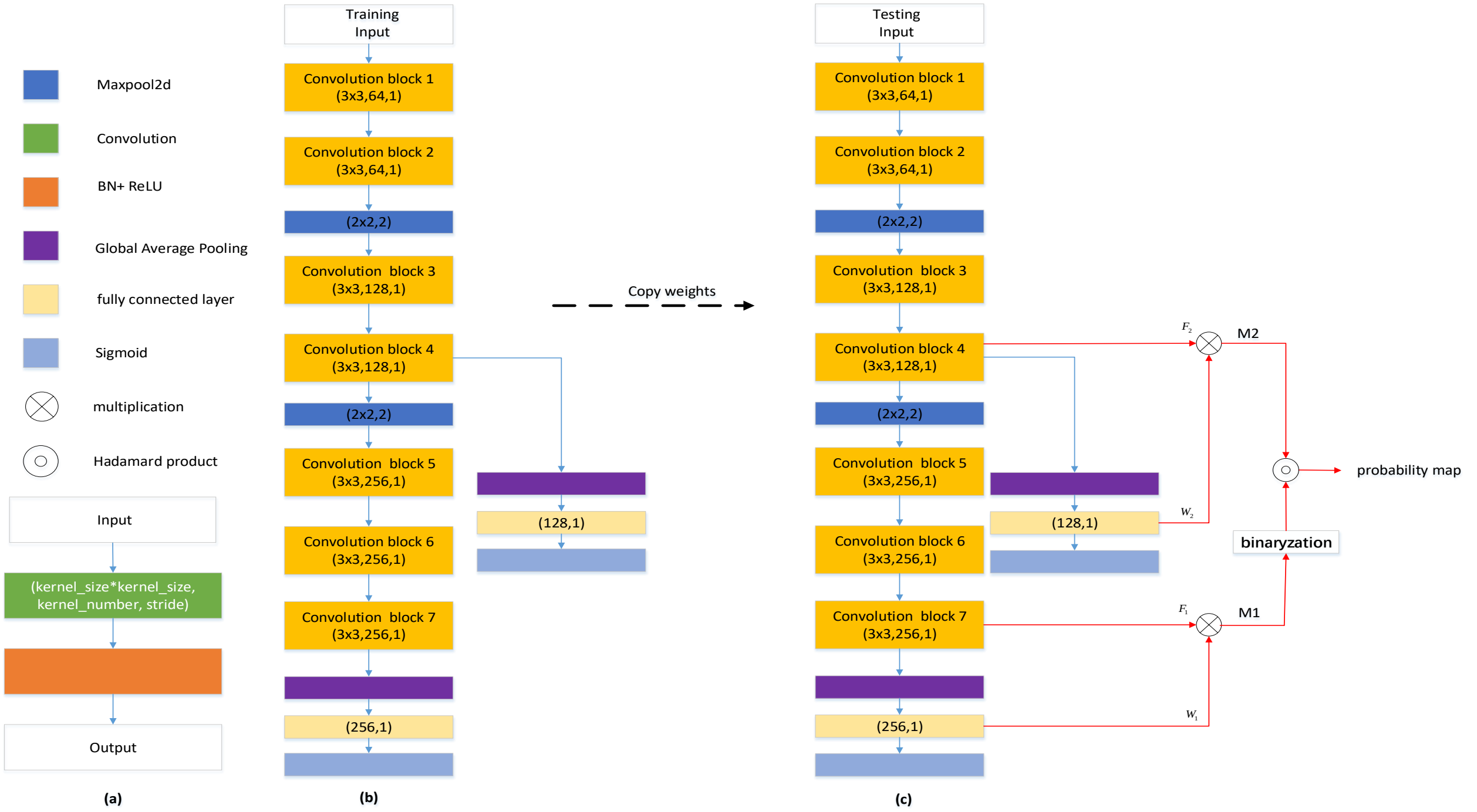

Fig. 5. Our proposed method for generating probability maps. (a) Convolution block; (b) Double-Path Classification network; (c) Inference process of the proposed DPC-Net. Best viewed in color.

### 3.1 Double-Path Classification Network

The DPC-Net is proposed based on the VGG-16 network [37] truncated before the third maxpooling layer, and the network architecture is depicted in Fig. 5(b). As we can see, we added a global average pooling (GAP) layer followed by a fully connected (FC) layer at the top of the network. Intuitively, the feature maps of the convolution block 7 is much lower than the original input images, leading to an inaccurate output probability map from the view of segmentation task. To this end, we further added a side-branch with a GAP layer and a FC layer on convolution block 4 to improve the spatial resolution. As we will see in Section 4, despite that the side-branch output may contain many false positives due to the lack of semantic information, the segmentation results can be significantly improved by combining the outputs of the main-branch and the side-branch.

At the training stage, as only the slice-level label is available, a classifier is trained by using the DPC-Net shown in Fig. 5(b), which can be regarded as a weakly-supervised leaning in the viewpoint of the AIS lesion segmentation. At the inference stage, as our objective is to locate the AIS lesions, the class activation maps (CAMs) [38] is computed for both the main and the side branches to generate the segmentation seeds with adequate semantic information to highlight suspicious lesion region.

In particular, as Fig. 5(c) shows, we generate the CAMs from conv-block 7 and conv-block 4, and then bilinearly upsample them to the same size as the original input, denoted as $\mathbf{M}_1$ and $\mathbf{M}_2$, respectively, where the weights are copied from the corresponding FC layers. To ensure that the maximum activation equals one, the CAMs are further normalized as

$$\bar{\mathbf{M}}_{\mathbf{i}} = \begin{cases} \mathbf{M}_{\mathbf{i}} / \max \mathbf{M}_{\mathbf{i}}, & \hat{y} \ge 0.5 \\ 0, & \hat{y} < 0.5 \end{cases} \tag{1}$$

for $i = 1, 2$, where $\hat{y}$ denotes the classifier output on the main branch.

Note that the main branch CAM output $\bar{\mathbf{M}}_1$ has a low spatial resolution with, however, more semantic information. We propose to fuse both CAMs to achieve a more accurate segmentation of the lesion regions. We first compute a binary segmentation map $\bar{\mathbf{M}}_{\mathbf{1b}}$ from the main branch CAM output $\bar{\mathbf{M}}_1$ by a threshold of 0.5, and then fuse $\bar{\mathbf{M}}_{\mathbf{1b}}$ and $\bar{\mathbf{M}}_2$ to generate PM as

$$\mathbf{M}_{\mathbf{p}} = \bar{\mathbf{M}}_{\mathbf{1b}} \circ \bar{\mathbf{M}}_2 \tag{2}$$

where $\circ$ denotes the Hadamard product.

### 3.2 Segmentation Map Generation

The K-Means is an unsupervised machine-learning algorithm that partition a set of vectors into $K$ groups that cluster around a common mean vector. When applying to a single image, it can be interpreted as a segmentation method that clusters image pixels with similar intensities to the same groups, making it accurate in segmenting the lesions with distinguished intensity property. On the other hand, different from the deep-learning-based segmentation, the K-Means clustering focus only on the pixel intensities, and no semantic information is considered, leading to many false positives in lesion segmentation.

In this paper, by making use of the a priori knowledge that the AIS lesions appear as hyperintense on the DWIs [3], we propose to adopt K-Means algorithm and a region-growing algorithm to identify the lesion regions. In particular, the initial growing points of region-growing algorithm are given by the outputs of DPC-Net rather than manual points. Our proposed algorithm is summarized as Algorithm 1.

Algorithm 1 integrates both the pixel-level information of the original DWIs and the semantic information extracted by the DPC-Net. To achieve a better performance, the key problem is how to determine the number of clusters $K$ and the threshold $\delta$. In this paper, we propose to use some fully-labeled subjects to fine-tune these two parameters by using grid search method. As we will discuss in the next section, the segmentation results can be significantly improved by using a very small amount of fully-labeled images.

**Algorithm 1** Segmentation Map Generation

**Input:** K-Means parameter $K$, DPC-Net output $\mathbf{M_p}$, threshold $\delta$

**Initialize:** AIS lesion segmentation map $\mathbf{Q} = \mathbf{0}$

1: Do K-Means on the DWI and cluster the pixel into $K$ groups.
2: Sort the $K$ groups by pixel intensity.
3: Preserve the group with the largest pixel values, and obtain a clustering map $\mathbf{I}$.
4: Do connected component analysis on $\mathbf{I}$, and identify all connected regions, denoted as $L_j$, for $j = 1, 2, \cdots$.
5: Select a point $x$ with $\mathbf{M_p} \geq \delta$.
6: Set $\mathbf{Q}(y) = 1, \forall y \in L_j$, if $x \in L_j$.
7: Stop if all points that satisfy $\mathbf{M_p} \geq \delta$ have been checked. Otherwise, return to Step 5.

## 3.3 Evaluation Metrics

In this paper, we propose to use dice coefficient (DC) to evaluate the pixel-level segmentation performance, which is defined as

$$\mathrm{DC} = \frac{2|G \cap P|}{|G| + |P|} \tag{3}$$

Where $G$ and $P$ denote the ground truth and the predicted segmentation, respectively. $|\bullet|$ denotes the region of lesion segmentation.

In clinical diagnosis, the segmentation results on both large and small lesions are of equal importance. It is therefore necessary to use the lesion-wise metrics to evaluate the performance. In particular, we perform 3D connected component analysis on both the ground truth and the predicted segmentation. A region is said to be a true positive (TP) if it appears on both ground truth and the prediction. A false positive (FP) is counted if a region on the prediction has no overlapping area with any region on the ground truth; while a false negative (FN) is counted if a region appears on the ground truth has no overlapping area with any region on the prediction. The mean number of TPs (m#TP), the mean number of FPs (m#FP) and the mean number of FNs (m#FN) can be then calculated by averaging over the total number of subjects.

We propose to use the lesion-wise precision rate ($\mathrm{P_L}$), the lesion-wise recall rate ($\mathrm{R_L}$) and the lesion-wise $\mathrm{F_1}$ score as lesion-wise metrics, which are given as

$$\mathrm{P_L} = \frac{\mathrm{m\#TP}}{\mathrm{m\#TP} + \mathrm{m\#FP}} \tag{4}$$

$$R_\mathrm{L} = \frac{\mathrm{m\#TP}}{\mathrm{m\#TP} + \mathrm{m\#FN}} \tag{5}$$

and

$$\mathrm{F_1} = \frac{2\mathrm{P}_L \cdot \mathrm{R}_L}{\mathrm{P}_L + \mathrm{R}_L} \tag{6}$$

respectively.

# 4. Experiment and Results

## 4.1 Data and Preprocessing

The experimental data used in this study were collected from Tianjin Huanhu Hospital, which includes 615 patients with AIS lesions. All clinical images were collected from a retrospective database and anonymized prior to use. Ethical approval was granted by Tianjin Huanhu Hospital Medical Ethics Committee. MR images were acquired from three MR scanners, with two 3T MR scanners (Skyra, Siemens and Trio, Simens) and one 1.5T MR scanner (Avanto, Siemens). DWI images were acquired using a spin-echo type echo-planar (SE-EPI) sequence with $b$ values of 0 and 1000 s/mm$^2$. The parameters are summarized in Table 1. ADC maps were calculated from the scan raw data in a pixel-by-pixel manner as

$$\mathrm{ADC} = \frac{\ln s_1 - \ln s_0}{b_1 - b_0} \tag{7}$$

where $b$ is the diffusion-sensitizing gradient pulses, with $b_1 = 1000$ s/mm$^2$ and $b_0 = 0$ s/mm$^2$ in our data. $s_1$ is the diffusion-weighted signal intensity with $b = 1000$ s/mm$^2$. $s_0$ is the signal with no diffusion gradient applied, i.e., with $b = 0$ s/mm$^2$.

The DWI and the corresponding ADC map were copy referenced to ensure the same slice position so as to allow optimal image evaluation and measurement. The ischemic lesions were manually annotated by two experienced experts (Dr. Song Jin and Dr. Chen Cao) from Tianjin Huanhu Hospital. We split the whole dataset into two subsets: training set and test set. The training set includes 460 subjects with weak labels for training and validating the DPC-Net, and 5 subjects with precise pixel-level labels to fine-tune clustering number $K$ in the K-Means algorithm and the threshold value $\delta$ in the region growing algorithm. The test set includes 150 subjects with full annotations to evaluate the segmentation performance.

Table 1
Parameters used in DWI acquisition.

| MR scanners | Skyra | Trio | Avanto |
|---|---|---|---|
| Repetition time (ms) | 5200 | 3100 | 3800 |
| Echo time (ms) | 80 | 99 | 102 |
| Flip angle (o) | 150 | 120 | 150 |
| Number of excitations | 1 | 1 | 3 |
| Field of view (mm2) | 240 × 240 | 200 × 200 | 240 × 240 |
| Matrix size | 130 × 130 | 132 × 132 | 192 × 192 |
| Slice thickness (mm) | 5 | 6 | 5 |
| Slice spacing (mm) | 1.5 | 1.8 | 1.5 |
| Number of slices | 21 | 17 | 21 |

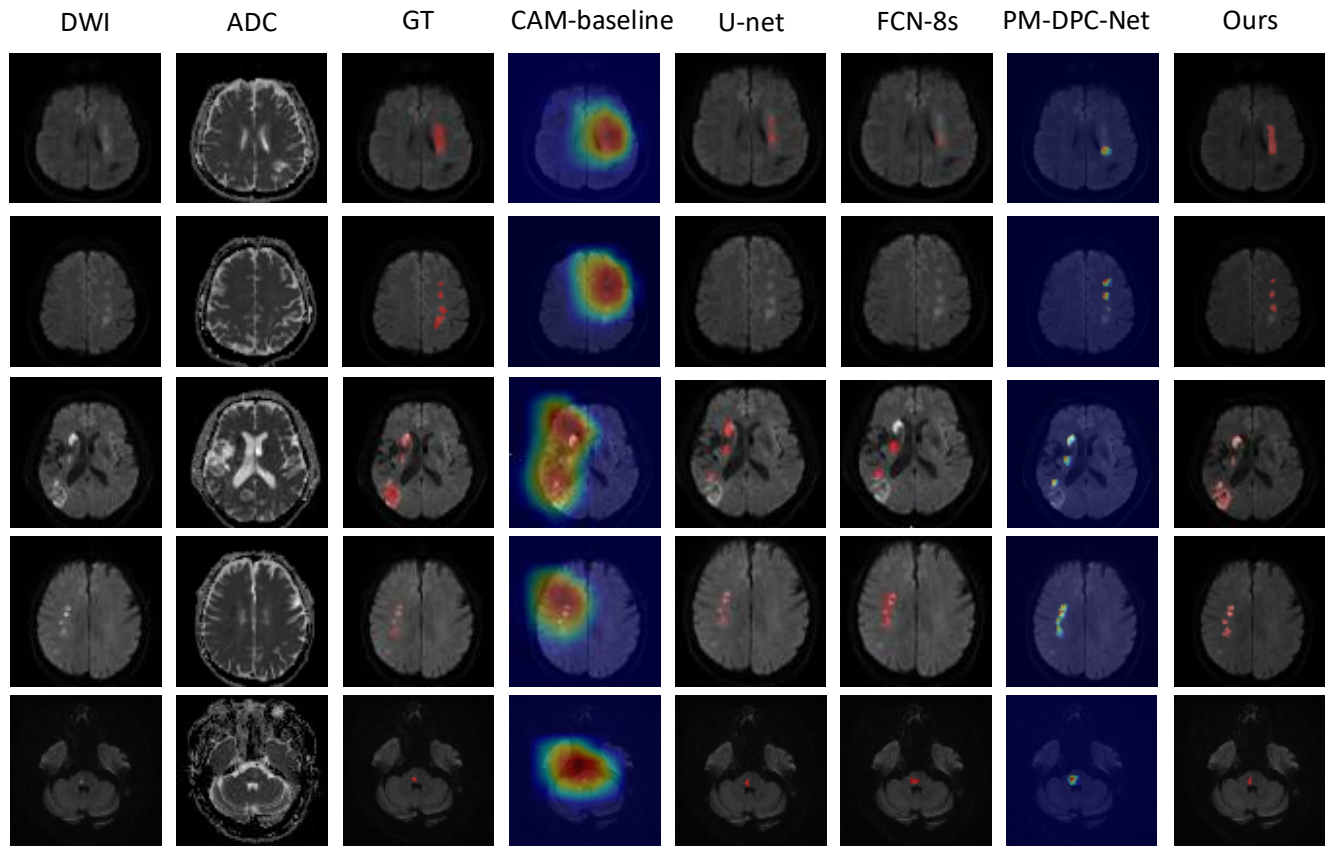

Fig. 6. Examples of lesion segmentation, CAMs and PMs. The first two columns show the original DWI and ADC map, respectively. The third and the eighth columns show the manually labeled lesions and the segmentation results of our semi-supervised method, respectively. The fourth to the sixth column shows the CAMs of the baseline VGG-16 Network (CAM-baseline [38]), the segmentation results of U-net [39]and FCN-8s [20], and the seventh column shows the PMs of DPC-Net (PM-DPC-Net). The CAMs and PMs are depicted on the DWI, and the redder with the color map, the more likely it is to represent the lesion region. The segmentations are also depicted on the DWI, and highlighted in red. Best viewed in color.

As the MR images were acquired on the three different MR scanners, the matrix size varies. Therefore, we resample all the MR images to the same size of 192 × 192 using linear interpolation. The pixel intensity of each MR image slice is normalized into that of zero mean and unit variance, and the DWI and ADC slices are concatenated into a dual-channel images and fed into the DPC-Net. During training, data augmentation technique is adopted to avoid over-fitting. In particular, each input image is randomly rotated by a degree ranging from 1 to 360 degree, flipped vertically and horizontally on the fly, so as to augment the dataset and reduce memory footprint.

## 4.2 Setup and Implementation

The hyperparameters of the proposed DPC-Net are shown in Fig. 5. We initialize these networks by Xavier's method [40] and use the Adam method [41] with $\beta_1 = 0.9$ , $\beta_2 = 0.999$ and initial learning rate of 0.001 as our optimizer. The learning rate is scaled down by a factor of 0.1 if no progress is made for 15 epochs on validation loss. Early-stopping technique is adopted after 30 epochs with no progress on the validation loss.

The experiments are performed on a computer with an Intel Core i7-6800K CPU, 64GB RAM and Nvidia GeForce 1080Ti GPU with 11GB memory. The computer operates on Windows 10 with CUDA 9.0. The network is implemented on PyTorch 0.4(https://pytorch.org/) and K-Means is implemented on scikit-learn (https://scikit-learn.org/stable). The MR image files are stored as Neuroimaging Informatics Technology Initiative (NIfTI) format, and processed using Simple Insight ToolKit (SimpleITK) [42]. We use ITK-SNAP [43] for the visualization of results.

## 4.3 Results

In our experiment, 460 weakly-labeled subjects and 5 fully-labeled subjects are used to train and fine-tune the parameters. The 5 fully-labeled subjects are abbreviated as fine-tuning set. Fig. 6 presents the segmentation results of our proposed method. The CAM results obtained from a naïve VGG-16 based CAM method [38], denoted as CAM-baseline, and the output probability map of our proposed DPC-Net, denoted as PM-DPC-Net, are also presented for comparison. As shown in Fig. 6, the CAM-based methods can successfully identify and localize the AIS lesions. Despite that the magnetic artifacts have similar appearances as the AIS lesions on the DWIs, the deep-learning-based methods can distinguish the lesions from the artifacts thanks to the semantic information extracted from the weakly-labeled subjects. From the aspect of segmentation, however, the CAM-baseline tends to segment much larger area than the actual AIS lesion, due to the fact that the output probability maps in conventional CAM method is obtained from features maps with much lower resolution than the original images. Our proposed DPC-Net significantly reduces the areas of the suspicious regions thanks to the side branch output, but underestimates the lesion areas in some cases. By integrating both pixel-level clustering and the PM-DPC-Net output, the proposed method is sensitive to both large and small lesions, and presents much better segmentation results over the CAM-baseline and PM-DPC-Net.

For the sake of comparison with other methods of semantic segmentation, we also train and evaluate on U-net [39] and FCN-8s [20] with our data set. We transfer the pretrained parameters of the DPC-Net to the FCN-8s and the encoder of U-net for fairness, and train these two networks with 5 fully-labeled subjects. As the Fig. 6 shows, the U-net and FCN-8s ignore some lesions in segmentation process for the reason that they are trained inadequately with 5 fully-labeled subjects. Besides, FCN-8s uses three-scale feature fusion and the outputs of its last convolutional layer resampled to the size of input image require interpolation of 32 times, which will overestimate the lesion region.

Table 2
The evaluation measurements on testing set. The best results are highlighted in bold.

| Method | ( $\delta$ , $K$ ) | $DC$ | $P_L$ | $R_L$ | $F1$ |
|---|---|---|---|---|---|
| CAM-baseline[38] | (0.7, ∗) | 0.091 | 0.854 | **0.791** | 0.821 |
| U-net[39] | (∗, ∗) | 0.582 | 0.791 | 0.670 | 0.726 |
| FCN-8s[20] | (∗, ∗ ) | 0.405 | **0.953** | 0.622 | 0.753 |
| PM-DPC-Net | (0.3, ∗) | 0.475 | 0.790 | 0.742 | 0.765 |
| Ours | (0.41,6) | **0.642** | 0.880 | 0.772 | **0.822** |

∗ indicates that the method does not have this parameter $K$ or $\delta$ .

Table 3
Evaluation results under the variety of $K$ and $\delta$ on fine-tuning set (FT-set) and on test set.

| **($\delta$, $K$)** | (0.7, 4) | | (0.6, 5) | | (0.4, 6) | | (0.3, 7) | |
|---|---|---|---|---|---|---|---|---|
| Data set | FT-set | Test set | FT-set | Test set | FT-set | Test set | FT-set | Test set |
| $\boldsymbol{DC}$ | 0.198 | 0.322 | 0.411 | 0.594 | 0.556 | 0.642 | 0.366 | 0.610 |
| $\boldsymbol{P_L}$ | 0.857 | 0.951 | 0.750 | 0.919 | 0.778 | 0.880 | 0.778 | 0.890 |
| $\boldsymbol{R_L}$ | 0.750 | 0.789 | 0.750 | 0.769 | 0.875 | 0.772 | 0.875 | 0.741 |
| $\boldsymbol{F1}$ | 0.800 | 0.862 | 0.750 | 0.837 | 0.824 | 0.822 | 0.824 | 0.809 |

The numerical evaluation results on the test set with 150 fully-labeled subjects are summarized in Table 2. For CAM-baseline and PM-DPC-Net, the thresholds to generate the binary segmentation are selected by using the fine-tuning set. For our proposed method, these fully-labeled images are used to tune both the threshold $\delta$ and the number of clusters $K$. As Table 2 shows, our proposed method achieves the best results on test set. From the aspect of the pixel-level metrics, our proposed method achieves a DC of 0.642, which is high than the results obtained by U-net, FCN-8s and PM-DPC-Net, and substantially exceeds the results obtained by CAM-baseline. In fact, such performance is very close to the method trained on fully-labeled images [24]. From the aspect of lesion-wise metrics, our proposed method achieves the precision rate of 0.880, which is higher than other methods except for FCN-8s. In fact, FCN-8s has 19 missed diagnosis subjects while our method only has 2 missed diagnosis, which leads the less FPs of FCN-8s. The recall rate, however, is slightly worse than the CAM-baseline due to the fact that the CAM-baseline tends to annotate a very large area. The K-Means clustering algorithm may also fail to group all lesions to the same group if the intensities of different lesions differ significantly from each other.

## 5. Discussions

### 5.1 Effect of Clustering Numbers

As the AIS lesions appear as hyperintense on the DWI, we adopt K-Means clustering algorithm to identify the hyperintensive regions. Note that the artifacts on the DWIs are also the hyperintensive regions, making it crucial to fine tune the value of $K$. As we can see from Fig. 7, when $K = 4$, more artifacts will be in the same cluster as the AIS lesions including the artifacts around the lesions, thus, the clustering lesion regions will be larger than the true lesions. As the $K$ increases, the clustering lesion regions will gradually decrease from greater than the true lesion regions to the near true lesion regions, and then to less than the true lesion regions until some clustering lesion regions disappear. The first row and the fourth row in Fig. 7 show that some clustering lesion regions have disappeared when $K = 7$.

In our work, we propose to use a small amount of fully-labeled subjects to search the optimal parameters in a grid search manner, and use DC as the metric. The results with $K = 4,5,6,7$ are summarized in Table 3. As we can see from Table 3, the best threshold $\delta$ increases as $K$ when $K = 4,5,6$, the performance is also improved. However, when $K = 7$, the threshold $\delta$ reduces to 0.3. This phenomenon indicates that when the clustering lesion regions reduce to the near true lesion regions, the predicted suspicious lesion regions in corresponding PM should have small intersections with the clustering lesion regions in clustering maps. However, when some clustering lesion regions disappear, the predicted suspicious lesion regions in corresponding PM will expand its lesion regions. The optimal clustering number leads to the best clustering results in general.

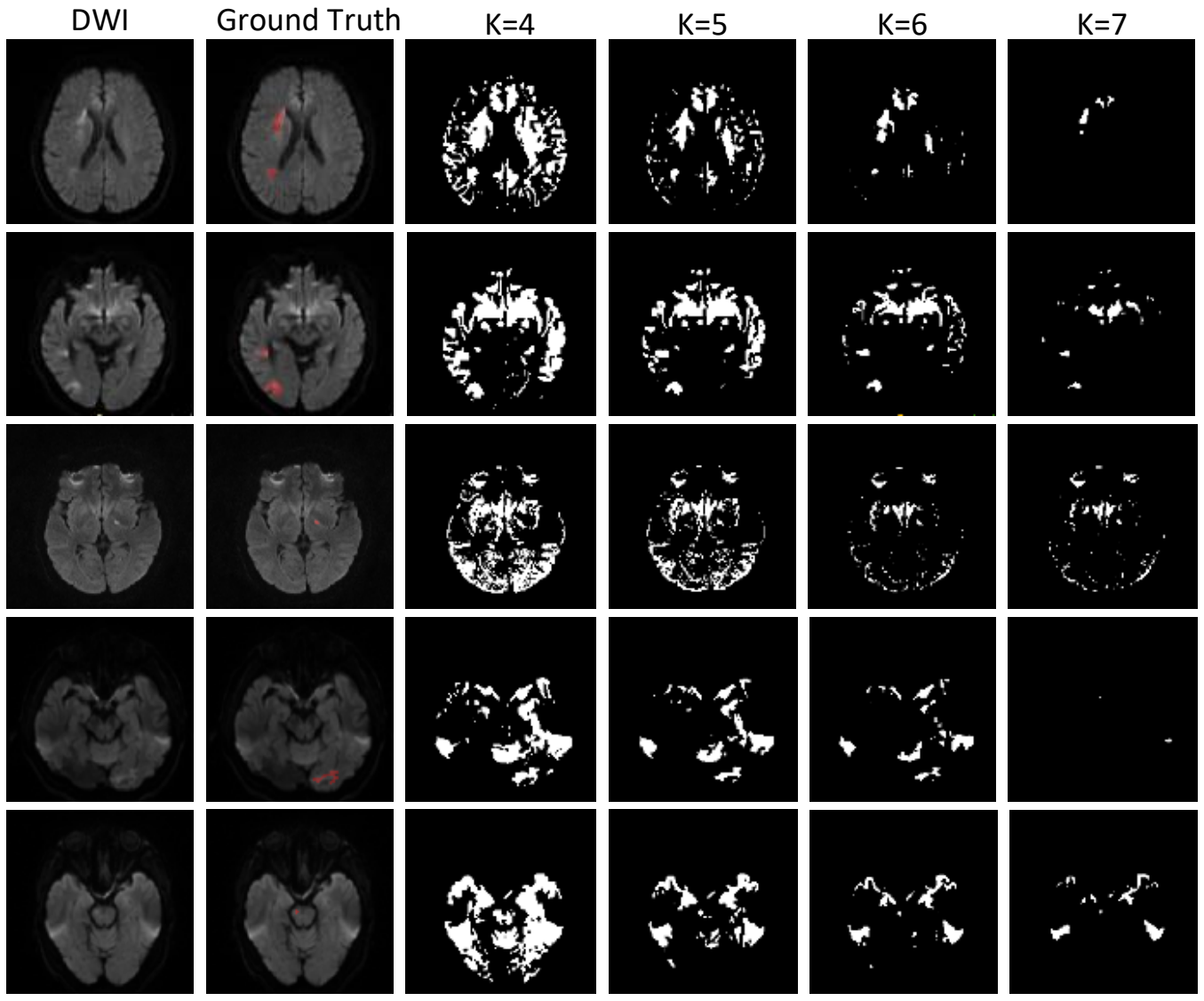

Fig. 7. Examples of clustering map. The first two columns show the original DWI and the manually labeled lesions, respectively. The last four columns show the clustering maps with the clustering number of 4,5,6 and 7, respectively.

### 5.2 Impact of Lesion Size

As pointed out in [44], a AIS lesion is classified as a lacunar infarction (LI) lesion if its diameter is smaller than 1.5cm. Clinically, the LI stroke accounts for 85% of all AIS patients. However, it is much difficult to be diagnosed in clinical practice, especially when it is too small to be noticed. It is, therefore, very necessary to evaluate the performance on subjects with small lesions.

In this subsection, we further divide our test sets to a large lesion set and a small lesion set. A subject is categorized into the small lesion set only if all of the lesions are LI lesions. Otherwise, it will be included in the large lesion set. In our test set, the large lesion and the small lesion sets include 60 subjects and 90 subjects respectively. As Table 4 shows, our proposed method achieves a DC of 0.708 on the small lesion set, while a DC of 0.543 on the large lesion set. Meanwhile, the $R_L$ of small lesions is higher than that of large lesions. The reason is that the distribution of hyperintense in large lesion regions are uneven, which makes the lesion regions predicted by DPC-Net to be smaller than the true lesion regions. Thus, how to solve the influence of hyperintense distribution imbalance on lesion segmentation is a problem to be considered in the future.

In clinical diagnosis, large lesions are more easily diagnosed, while small lesions are not. Our proposed method achieves high performance on small lesions, which might be of a good inspiration for other methods.

Table 4
Evaluation results on test set, large lesions set and small lesions set, respectively. The best results are highlighted in bold.

| Data set | $DC$ | $P_L$ | $R_L$ | $F1$ |
|---|---|---|---|---|
| Test set | 0.642 | 0.880 | 0.772 | 0.822 |
| Large lesions set | 0.543 | **0.889** | 0.705 | 0.786 |
| Small lesions set | **0.708** | 0.867 | **0.901** | **0.883** |

## 6. Conclusion

In this paper, we present a semi-supervised method for AIS lesion segmentation, where 460 weakly-labeled subjects are used to train the DPC-Net and then 5 fully-labeled subjects are used to fine-tune the parameters in a supervised way.

The proposed semi-supervised method presents a high segmentation accuracy on the clinical MR images with a dice coefficient of 0.642. More importantly, it presents very high precision of 0.880, which is of paramount importance in avoiding misdiagnosis in clinical scenario. Meanwhile, the proposed method largely reduces the expense of obtaining a large number of fully-labeled subjects in a supervised setting, which is more meaningful in terms of engineering maneuverability.

## Conflict of interest

All authors of this manuscript do not have financial and personal relationships with other people or organizations that could inappropriately influence their work.

## Acknowledgements

This work is supported in part by the National Natural Science Foundation of China [grant numbers 61871239, 61571244 and 61671254], and in part by the Fundamental Research Funds for the Central Universities, Nankai University [grant number 63191106].